\documentclass[journal,10pt,twocolumn]{IEEEtran}

\usepackage{threeparttable}
\usepackage{cite}
\usepackage{amsmath,amssymb,amsfonts}
\usepackage[ruled,vlined,linesnumbered,noresetcount]{algorithm2e}
\usepackage{graphicx}
\usepackage{textcomp}
\usepackage{indentfirst}
\usepackage{csquotes}
\usepackage{multirow}
\usepackage{float}
\usepackage{color}
\usepackage{comment}
\usepackage{caption2}
\usepackage{braket}
\usepackage{url}

\newcommand{\myalgorithmname}{SABRE}
\begin{document}

\title{Tackling the Qubit Mapping Problem for NISQ-Era Quantum Devices}

\author{Gushu Li, Yufei Ding, and Yuan Xie \\
Unversity of California, Santa Barbara, CA, 93106, USA   \\ 
\{gushuli,yufeiding,yuanxie\}@ucsb.edu}

\maketitle

\begin{abstract}
Due to little consideration in the hardware constraints, e.g., limited connections between physical qubits to enable two-qubit gates, most quantum algorithms cannot be directly executed on the Noisy Intermediate-Scale Quantum~(NISQ) devices. 
Dynamically remapping logical qubits to physical qubits in the compiler is needed to enable the two-qubit gates in the algorithm, which introduces additional operations and inevitably reduces the fidelity of the algorithm. Previous solutions in finding such remapping suffer from high complexity, poor initial mapping quality, and limited flexibility and controllability.

To address these drawbacks mentioned above, this paper proposes a \textbf{S}W\textbf{A}P-based \textbf{B}idi\textbf{RE}ctional heuristic search algorithm (\myalgorithmname), which is applicable to NISQ devices with arbitrary connections between qubits. By optimizing every search attempt,
globally optimizing the initial mapping using a novel reverse traversal technique, introducing the decay effect to enable the trade-off between the depth and the number of gates of the entire algorithm, \myalgorithmname~outperforms the best known algorithm with exponential speedup and comparable or better results on various benchmarks.


\end{abstract}

\section{Introduction}
Quantum Computing~(QC) has been rapidly growing in the last decades because of its 
potential in various important applications, including integer factorization~\cite{shor1999polynomial}, database search~\cite{grover1996fast}, quantum simulation~\cite{peruzzo2014variational}, etc. 
Recently, IBM, Intel, and Google released their QC devices with 50, 49, and 72 qubits respectively~\cite{IBM50Q, Intel49Q, Google72Q}. 
IBM and Rigetti also provide cloud QC services~\cite{IBMdevice,RigettiQPU}, allowing more people to study real quantum hardware. 
We are expected to enter the Noisy Intermediate-Scale Quantum~(NISQ) era in the next few years~\cite{preskill2018quantum}, when QC devices with dozens to hundreds of qubits will be available. 
Though the number of qubits is insufficient for Quantum Error Correction~(QEC), it is expected to use these devices to solve real-world problems beyond the capability of available classical computers~\cite{preskill2012quantum,boixo2018characterizing}.

However, there exists a gap between quantum software and hardware due to technology constraints in the NISQ era. 
When designing a quantum program based on the most popular 
circuit model, it is always assumed that qubits and quantum operations are perfect and you can apply any quantum-physics-allowed operations.
But on NISQ hardware, the qubits have limited coherence time and quantum operations are not perfect. 
Furthermore, only a subset of theoretically possible quantum operations can be directly implemented, which calls for a modification in the quantum program to fit the target platform.

In this paper, we will focus on the \textit{qubit mapping problem} caused by limited two-qubit coupling on NISQ devices. 
Two-qubit gates are one important type of quantum operations applied on two qubits. 
They can create quantum entanglement, an advantage that does not exist in classical computing.
Two-qubit gates can be applied to arbitrary two logical qubits in a quantum algorithm but this assumption does not hold with NISQ devices.
When running a quantum program, the logical qubits in the quantum circuit need to be mapped to the physical qubits (an analogy in classical computation is register allocation). 
But for the physical qubits on NISQ devices, one qubit can only couple with its neighbor qubits directly. 
So that for a specific mapping, two-qubit gates can only be applied to limited logical qubit pairs, whose corresponding physical qubit pairs support direct coupling.
This makes a quantum circuit not directly executable on NISQ devices.

As a result, circuit transformation is required to make the circuit compatible with NISQ device during compilation. 
Based on a given quantum circuit and the coupling information of the device, we need 1) an initial logical-to-physical qubit mapping and 2) the intermediate mapping transition which is able to remap the two logical qubits in a two-qubit gate to two coupled physical qubits.
The qubit mapping problem has been proved to be NP-Complete~\cite{siraichi2018qubit}.

Previous solutions to this problem can be classified into two types. One type is to formulate it into an equivalent mathematical problem and then apply a solver~\cite{maslov2008quantum,chakrabarti2011linear,shafaei2013optimization,shafaei2014qubit,wille2014optimal,lye2015determining,venturelli2017temporal,venturelli2018compiling,booth2018comparing,oddi2018greedy,bhattacharjee2017depth}. 
These attempts suffer from very long runtime and can only be applied to small size cases. 
Moreover, general software solvers can not exploit the intrinsic feature of the quantum mapping problem.  
Another type of approach is heuristic search~\cite{alfailakawi2014lnn,saeedi2011synthesis,lin2015paqcs,wille2016look,shrivastwa2015fast,kole2016heuristic,kole2018new,bhattacharjee2018novel}, while most of them were developed on ideal 1D/2D lattice model and not applicable to NISQ devices with more irregular and restricted coupling connections. 
Some recent works~\cite{IBMqiskit,siraichi2018qubit,zulehner2018efficient} targeting IBM chip architecture are able to handle arbitrary coupling but they suffer from very long runtime due to exhaustive mapping search and their solutions for initial mapping lack the ability of global optimization. Moreover, none of them has the ability to control the generated circuit quality among multiple optimization objectives to fit in NISQ devices with different characteristics.

In this paper, a \textbf{S}W\textbf{A}P-based \textbf{B}idi\textbf{RE}ctional heuristic search algorithm, named \myalgorithmname, is proposed to solve this qubit mapping problem and overcome the drawbacks mentioned above.
With the observation that many attempts in exhaustive search can be redundant and effective mapping transition needs to start from the qubits in the two-qubit gates that need to be executed, 
we design an optimized SWAP-based heuristic search scheme in \myalgorithmname~with significantly reduced search space. 
Initial mapping has been proved to be very important in this problem, which can significantly affect the final circuit quality~\cite{siraichi2018qubit,zulehner2018efficient}.
We present a novel reserve traversal search technique in \myalgorithmname~to naturally generate a high-quality initial mapping through traversing a reverse circuit, in which more consideration is given to those gates at the beginning of the circuit without completely ignoring the rest of the circuit. 
Moreover, we introduce a $decay$ effect, which will slightly increase our heuristic cost function values when evaluating overlapped SWAPs, to let \myalgorithmname~tend to select non-overlapped SWAPs.
This optimization enables the control of parallelism in the additional SWAPs and can further generate different hardware-compliant circuits with a trade-off between circuit depth and the number of gates.

\myalgorithmname~is evaluated with various benchmarks on a latest IBM 20-qubit chip model~\cite{IBMdevice} compared with the best known solution~\cite{zulehner2018efficient}. 
Experimental results show that \myalgorithmname~is able to find the optimal mapping for small benchmarks.
The number of additional gates is reduced by 91\% or even fully eliminated.
For larger benchmarks, \myalgorithmname~can demonstrate exponential speedup against the previous solution and still outperform it by about 10\% reduction in the number of additional gates on average with the assistance of the high-quality initial mapping generated by our proposed method.
In some cases, the best known previous solution cannot even finish execution due to exponential execution time and memory requirement while \myalgorithmname~can still work with short execution time and low memory usage. 
By tuning the decay parameters in our algorithm, \myalgorithmname~shows the ability to control the generated circuit quality with about 8\% variation in generated circuit depth by varying the number of gates.

The major contributions of this paper can be summarized as follows:
\begin{itemize}
\item We perform a comprehensive analysis on the shortcomings of previous solutions, and then summarize the objectives and metrics that should be considered when designing a heuristic solution for the qubit mapping problem. 

\item We propose a SWAP-based search scheme which can produce comparable results with an exponential speedup in the search complexity compared with previous exhaustive mapping-based search algorithms.
This fast search scheme ensures the scalability of \myalgorithmname~to accommodate larger-size quantum devices in the NISQ era. 

\item We present a reverse traversal technique to enable global optimization in the initial mapping solution by leveraging the intrinsic reversibility in qubit mapping problem. Our high-quality initial mapping can significantly reduce the overhead in the generated circuit. 

\item By introducing a decay effect in the heuristic cost function, we are able to generate different hardware-compliant circuits by trading the number of gates in the circuit against the circuit depth.
This makes \myalgorithmname~applicable for NISQ devices with different characteristics and optimization objectives.
\end{itemize}

The rest of this paper is organized as follows. We introduce QC background information in Section~\ref{sec:background} and then 
formulate the qubit mapping problem in Section~\ref{sec:problem}. 
Our solution \myalgorithmname~is introduced in Section~\ref{sec:methodology} and evaluated in Section~\ref{sec:evaluation}. 
Limitations and future research directions are discussed in Section~\ref{sec:futurework}.
Related works are summarized in Section~\ref{sec:relatedwork} and we finally conclude this paper in Section~\ref{sec:conclusion}.

\section{Background}\label{sec:background}
In this section, we will give a brief introduction to QC.
QC research spans all technology stacks from high-level theory, algorithm, to mid-level architecture and low-level physics~\cite{van2013blueprint,chong2017programming,fu2017experimental}.
We try to limit our discussion and only keep the necessary content to help formulate and understand this qubit mapping problem.

\subsection{QC Software Basics}
Among several existing QC theoretical models which are mathematically equivalent, we will focus on the most popular quantum circuit model.
We will start from the basic concepts, including quantum bit (qubit) and quantum operations, and then establish quantum programs, which can be represented using quantum circuits.

\textbf{Qubit.} Classical bit is the basic information unit which has two deterministic states, `0' and `1'. 
One qubit also has two basis states, usually denoted as $\ket{0}$ and $\ket{1}$.
Different from classical bit, one qubit can be the linear combination of the two basis states, which can be represented by 
$\ket{\Psi} = \alpha\ket{0} + \beta\ket{1}$, where $ \alpha, \beta \in \mathbb{C}$ and $|\alpha|^2 + |\beta|^2 = 1$. The state vector is $(\alpha, \beta)$.
Moreover, two or more qubits can be entangled. The state of a two-qubit system can be represented by $\ket{\Psi} = \alpha_{00}\ket{00} + \alpha_{01}\ket{01} +  \alpha_{10}\ket{10} + \alpha_{11}\ket{11}$, whose state vector is $(\alpha_{00},\alpha_{01},\alpha_{10},\alpha_{11})$.

\textbf{Quantum Operation.} There are two types of quantum operations. The first one is Quantum Gates, which are unitary operations applied on qubits to modify the qubit states. A single-qubit gate is applied on one qubit. For example, Hadamard gate~(denoted as $H$ in a quantum circuit) is a widely used single-qubit operation which can be represented by a $2 \times 2$ matrix. Control-NOT~(CNOT) gate is a two-qubit operation, applied on two qubits simultaneously. It will flip a target qubit based on a control qubit. The second type is Measurement. We measure one qubit and the result can be either  $\ket{0}$ or $\ket{1}$ with the probability based on the state vector.

\begin{figure}[!t]
\centering
\includegraphics[width=1.0\columnwidth]{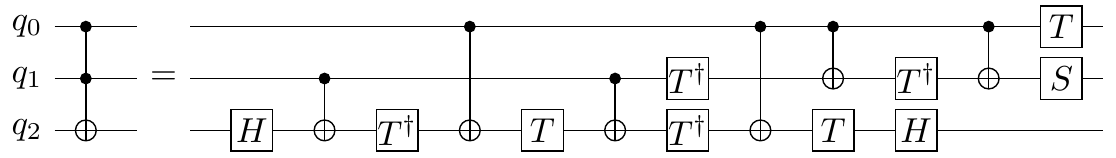}
\vspace{-15pt}
\caption{Example Quantum Circuit of Toffoli Gate}
\vspace{-5pt}
\label{fig:toffoli}
\end{figure}

\textbf{Quantum Circuit.} 
Quantum circuit is a diagram to represent a quantum program.
Each line in the quantum circuit represents one qubit and the operations are represented by different blocks on the line.
Figure~\ref{fig:toffoli} shows a quantum circuit that decomposes the Toffoli gate~\cite{nielsen2010quantum} using only single- and two-qubit gates. 
The three-qubit gate on the left is Toffoli gate. It can be decomposed into a gate sequence on the right side. One square represents a single-qubit gate and a line connecting two qubits represents a CNOT gate.
Barenco \textit{et al.} proved that arbitrary quantum circuit can be expressed by compositions of a set of single-qubit gates and CNOT gate~\cite{barenco1995elementary}. 
As a result, we only use single-qubit and CNOT gates, which also compose the elementary gate set directly supported by IBM quantum chips on cloud service, to construct quantum circuits in this paper.

\subsection{QC Hardware in the NISQ Era}\label{sec:hardware}
There are several different candidate technologies to implement QC on hardware, including superconducting quantum circuit~\cite{koch2007charge}, ion trap~\cite{nigg2014quantum}, quantum dot~\cite{zajac2016scalable}, neutral atom~\cite{saffman2010quantum}, etc.
We will use superconducting quantum circuits, which is currently the most promising technology, as an example to introduce QC hardware model.

\begin{figure}[!h]
\centering
\vspace{-5pt}
\includegraphics[width=1.0\columnwidth]{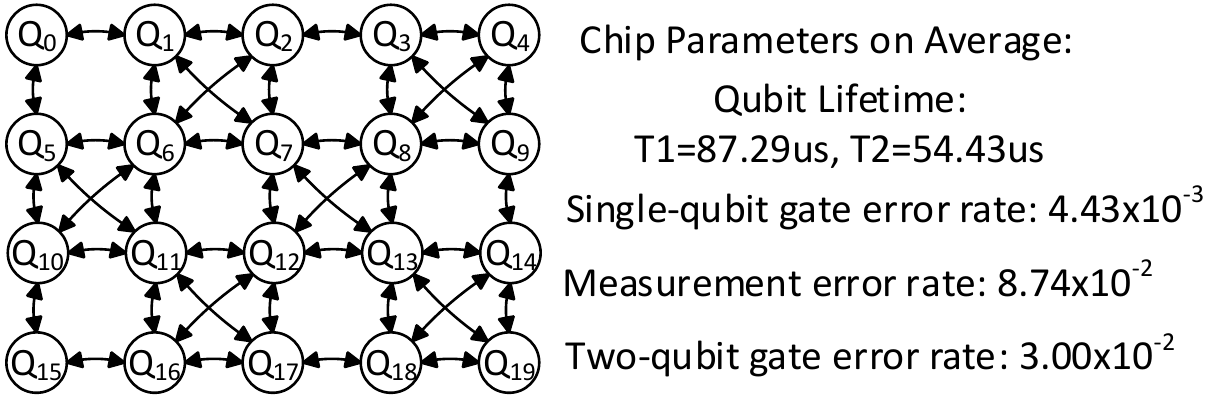}
\vspace{-13pt}
\caption{IBM Q20 Tokyo Information~\cite{IBMdevice}~(Vary over Time)}
\vspace{-5pt}
\label{fig:ibmq20}
\end{figure}




Figure~\ref{fig:ibmq20} shows the information about IBM Q20 chip~\cite{IBMdevice}. 
The lifetime of the qubits is about $50\mu\mathrm{s}$ on average.
The average error rates are $4.43\times 10^{-3}$, $8.47\times 10^{-2}$, $3.00\times 10^{-2}$ for single-qubit gate, measurement, and CNOT gate respectively. 
The coupling graph is shown on the left.
Two coupled qubits are connected by a bidirectional arrow.
The qubits are placed on a planar geometry and couplers can only connect one qubit to its neighboring qubits due to on-chip placement-and-routing constraints.
For example, $Q_0$ is connected to $Q_1$ and $Q_5$ through couplers, which means a CNOT gate can be applied on qubit pair $\{Q_0,Q_1\}$ and $\{Q_0,Q_5\}$ in either direction. However, $Q_0$ is not directly connected with $Q_6$ and you cannot apply a CNOT gate on these two qubits directly.


John Preskill proposed this NISQ concept, referring to quantum computers with the number of qubits ranging from dozens to hundreds~\cite{preskill2018quantum}. 
Quantum computers of such size are expected to appear in the next few years. 
Due to limited number of qubits in the NISQ era, all logical qubits in the quantum circuit are directly implemented by physical qubits without QEC.
NISQ hardware is not as perfect as the model used when we design a quantum program.
In this paper, the following three major limitations are considered:

\begin{enumerate}
\item \textbf{Qubit Lifetime.}
A qubit can only retain its state for a very short time. 
It may decay to another state or interact with the environment and lose the original quantum state.
The coherence time of state-of-the-art superconducting qubits can reach $\sim100~\mu\mathrm{s}$~\cite{IBMdevice}.
All the computation must be accomplished within a fraction of qubit coherence time, which sets an upper bound on the number of sequential gates that can be applied on qubits.

\item \textbf{Operation Fidelity.} 
Quantum operations applied to the qubits can also introduce errors. 
For example, the error rate for operations is reported to be around $10^{-3}$ for single-qubit gates, and $10^{-2}$ for two-qubit gates and measurements~\cite{IBMdevice,kelly2015state, walter2017rapid}.
Therefore, it is important to minimize the number of gates in a quantum algorithm to reduce the amount of error accumulated.

\item \textbf{Qubits Coupling.} 
A physical connection is required when applying two-qubit gates, which means that two-qubit gates can only be applied on two physically nearby qubits. One popular coupling structure is the 2D Nearest Neighbor structure which fits in the planar layout of qubits on state-of-the-art superconducting quantum chips.


\end{enumerate}


\section{Problem Analysis}\label{sec:problem}
In this section, we will illustrate the challenge of qubit mapping caused by the three limitations discussed above. We first introduce qubit mapping problem with a small-size example followed by a formal definition.
Then we will discuss the design objectives and the metrics we use to evaluate our solution.


\begin{figure*}[!t]
\centering
\includegraphics[width=2.0\columnwidth]{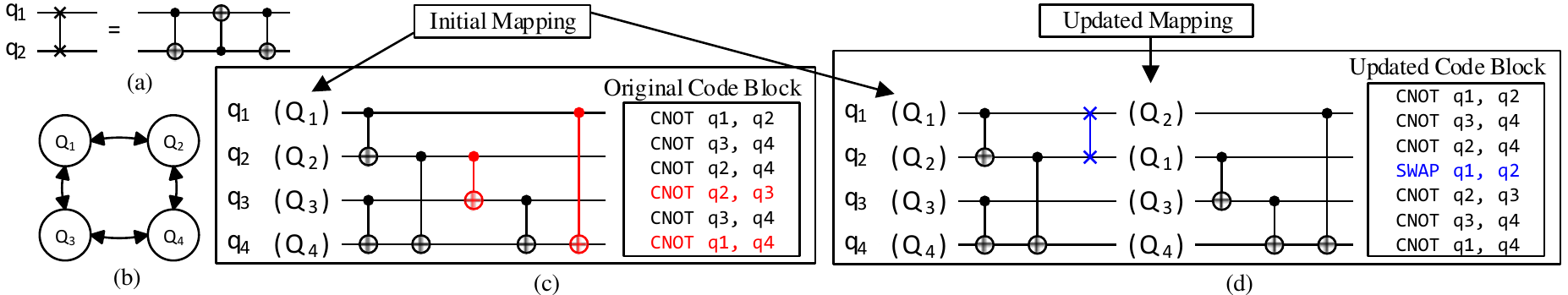}
\vspace{-5pt}
\captionsetup{justification=centering}
\caption{(a) SWAP Gate Decomposition, (b) Physical Qubit Coupling Graph Example, (c) Original Quantum Circuit, (d) Updated Hardware-Compliant Quantum Circuit}
\label{fig:swap_example}
\end{figure*}

\subsection{Problem in Qubit Mapping}
We use a small-size example to explain this qubit mapping problem. A 4-qubit device model is used as the hardware platform~( shown in Figure~\ref{fig:swap_example} (b)).
Two-qubit gates are allowed on the following physical qubit pairs:$\{Q_1,Q_2\}$, $\{Q_2, Q_4\}$, $\{Q_4, Q_3\}$, $\{Q_3, Q_1\}$ and not allowed on $\{Q_1, Q_4\}$, $\{Q_2, Q_3\}$.

Now suppose we have a small quantum circuit to be executed on this 4-qubit device. 
This quantum circuit consists of six CNOT gates~(shown in Figure~\ref{fig:swap_example} (c)).
We assume the initial logical-to-physical qubits mapping is $\{q_1\mapsto Q_1,q_2\mapsto Q_2,q_3\mapsto Q_3,q_4\mapsto Q_4\}$. 
We can find that four of the six CNOT gates can be directly executed but the fourth and the sixth CNOT gates~(marked red in Figure~\ref{fig:swap_example} (c)) cannot be executed because the corresponding qubit pairs are not connected on the device.
A perfect initial mapping to satisfy all two-qubit gate dependencies does not exist in this example and we need to change the qubit mapping during execution and make all CNOT gates executable.

\textbf{SWAP Qubit Mapping.} Same as previous solutions, we employ SWAP operations to change the qubit mapping by exchanging the states between two qubits. It consists of three CNOT gates~(shown in Figure~\ref{fig:swap_example} (a)). 
We can employ multiple SWAPs to move one logical qubit to arbitrary physical qubit location.
Even two qubits are not nearby on the quantum device, we can still move them together and then apply the two-qubit gate in the circuit.
Figure~\ref{fig:swap_example} (d) shows that the updated quantum circuit is now executable after we insert one SWAP operation between $q_1$ and $q_2$ after the third CNOT gates. The first three CNOT gates can be executed under initial mapping. After the inserted SWAP, mapping is updated to $\{q_1\mapsto Q_2,q_2\mapsto Q_1,q_3\mapsto Q_3,q_4\mapsto Q_4\}$. All three remaining CNOT gates now can be executed under this updated mapping.

\textbf{Other Methods.} Prior work also tried to employ other circuit transformation methods~\cite{siraichi2018qubit} like 'Reverse' or 'Bridge' because of the asymmetric connection hardware model from IBM's 5-qubit and 16-qubit chips~\cite{IBMdevice}. On those chips, CNOT gate is only allowed in one direction even two physical qubits are connected on the chip. 
Fortunately, physical experiments have shown that the connection between superconducting qubits can be symmetric~\cite{chen2014qubit} and on IBM's latest 20-qubit chip~\cite{IBMdevice,tannu2018case}, CNOT gate can already be applied on either direction between any connected qubit pair. 
Since the difficulty from the asymmetric connection is overcome by technology advance, we will focus on the latest symmetric coupling model and only consider inserting SWAPs for mapping change.

By introducing additional SWAPs in the quantum circuit, we can solve all the two-qubit gate dependencies and generate a hardware-compliant circuit without changing the original functionality. 
However, due to limitations of NISQ devices, inserting SWAPs in the quantum circuit will also cause the following problems:
\begin{enumerate}
\item The number of operations in the circuit is increased. Since the operations are imperfect and will introduce noise, the overall error rate will increase.
\item The depth of the circuit may also be increased, which means the total execution time will be increased and too much error can be accumulated due to qubit decoherence.
\end{enumerate}
If we compare the original circuit and the updated circuit in Figure~\ref{fig:swap_example} (c) and (d), the number of gates increases from 6 to 9 and the circuit depth increased from 5 to 8.
Additional SWAPs will bring significant overhead in terms of fidelity and execution time. 
As a result, we hope to minimize the number of additional SWAPs in order to reduce the overall error rate and total execution time for the final hardware-compliant circuit. 
We formally define the qubit mapping problem as follows:

\textbf{Definition:} Given an input quantum circuit and the coupling graph of a quantum device, find an \textbf{initial mapping} and the intermediate qubit \textbf{mapping transition} (by inserting SWAPs) to satisfy all two-qubit constraints and try to minimize the number of additional gates and circuit depth in the final hardware-compliant circuit.

\subsection{Objectives and Metrics}\label{sec:objective}

Since qubit mapping problem is NP-Complete~\cite{siraichi2018qubit}, it is hard to directly find the optimal solution.
We will design a heuristic algorithm trying to find a solution to this problem with the following objectives:

\begin{enumerate}
\item \textbf{Flexibility.}
NISQ devices may have an irregular coupling design which can evolve over time. Our algorithm should be able to deal with arbitrary symmetric coupling cases for various benchmarks.

\item \textbf{Fidelity.} 
This objective comes from the imperfect quantum operations. 
The error rate of a CNOT gate is high and one SWAP even requires 3 CNOT gates. 
We target to improve the overall fidelity by reducing the number of quantum gates, especially two-qubit gates, of the final hardware compliant circuit.

\item \textbf{Parallelism.} 
This objective comes from the limited qubit lifetime. Inserting SWAPs may increase the depth of the circuit. 
If our algorithm can insert SWAPs that can be executed in parallel and control the final circuit depth, we can allow a deeper circuit to be executed on hardware.

\item \textbf{Scalability.}
Our algorithm targets to be scalable with an acceptable execution time for NISQ devices which contain dozens to hundreds of qubits.
As the number of qubits continues to increase beyond the scope of NISQ in the future, QEC might be used, and the problem addressed in the paper turns into another one, as discussed in other papers~\cite{heckey2015compiler,paler2017fault,javadi2017optimized,lao2018mapping}.
\end{enumerate}

\textbf{Metrics.} 
Our algorithm is evaluated by a set of benchmarks of different sizes on IBM's latest public superconducting chip model~\cite{IBMdevice} to test the flexibility and scalability. The metrics of the evaluation are the total number of gates and the circuit depth in the final generated hardware-compliant circuit.


\section{Finding Initial Mapping and SWAPs}~\label{sec:methodology}
In this section, we will introduce our heuristic approach \myalgorithmname~step by step to illustrate how our design search could overcome the shortcomings of previous work. 
We start with preprocessing steps in Section~\ref{sec:preprocessing} and the overview of \myalgorithmname's SWAP-based heuristic search algorithm in Section~\ref{sec:flow}. Then we use several examples to explain key design decisions in \myalgorithmname~in Section~\ref{sec:decision}, followed by the heuristic function design in Section~\ref{sec:heuristic}.
We summarize the notations used in this paper in Table~\ref{tab:notation}.

\begin{table}[h!]
  \centering
  \caption{Definition of Notations used in this paper}
 \vspace{3pt}
  \begin{tabular}{|l|l|}
    \hline
    \textbf{Notation} & \textbf{Definition}\\
    \hline
    $n$ & number of logical qubits \\
    \hline
    $q_{\{1,2,\cdots,n\}}$ & logical qubits in quantum circuit \\
    \hline
    $g$ & number of gates in the circuit\\
    \hline
    $d$ & depth of the circuit\\
    \hline
    $N$ & number of physical qubits \\
    \hline
    $Q_{\{1,2,\cdots,N\}}$ & physical qubits on quantum device \\
    \hline
    $G(V,E)$ & the coupling graph of the chip\\
    \hline
    $D[~][~]$ &  the distance matrix of the physical qubits\\
            & $D[i][j]$ is the distance between $Q_i$,$Q_j$\\
    \hline
    $\pi()$ & a mapping from $q_{\{1,2,\cdots,n\}}$ to $Q_{\{1,2,\cdots,N\}}$ \\
    \hline
    $\pi^{-1}()$ & a mapping from $Q_{\{1,2,\cdots,N\}}$ to $q_{\{1,2,\cdots,n\}}$ \\
    \hline
    $F$ & Front Layer, defined in Section~\ref{sec:preprocessing} \\
    \hline
    $E$ & Extended Set, defined in Section~\ref{sec:heuristic} \\
    \hline
  \end{tabular}
  \label{tab:notation}
\end{table}

\subsection{Preprocessing}\label{sec:preprocessing}

Before we can begin our heuristic search, we have some preprocessing steps to prepare and initialize the required data.

\textbf{Distance matrix computing.} 
Given the coupling graph $G(V,E)$ of a quantum device, we will first compute the All-Pairs Shortest Path~(APSP) by Floyd-Warshall algorithm~\cite{floyd1962algorithm} to obtain the distance matrix $D[~][~]$.
Each edge in the coupling graph has distance 1 because one SWAP is required to exchange the two qubits of an edge. So that $D[i][j]$ represents the minimum number of SWAPs required to move a logical qubit from physical qubit $Q_i$ to $Q_j$.
The complexity of this step is $O(N^3)$, which is acceptable for NISQ devices with hundreds of qubits.

\begin{figure}[!t]
\centering
\includegraphics[width=1.0\columnwidth]{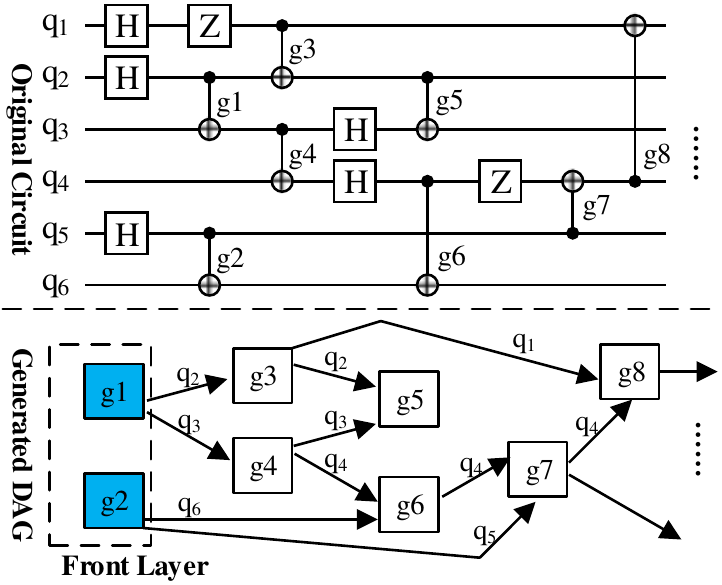}
\vspace{-18pt}
\caption{Example of DAG Generation and Front Layer Initialization.}
\vspace{-5pt}
\label{fig:DAG}
\end{figure}

\textbf{Circuit DAG generation.} 
We use a Directed Acyclic Graph (DAG) to represent the execution constraints between the two-qubit gates in a quantum circuit.
The single qubit gates are not considered here because they can always be executed locally on one qubit without bringing dependencies on other qubits.
A two-qubit gate $CNOT(q_i,q_j)$ can be executed only when all the previous two-qubit gates on $q_i$ or $q_j$ have been executed.
We traverse the entire quantum circuit and construct a DAG to represent execution dependencies with complexity $O(g)$.
An example is shown in Figure~\ref{fig:DAG}. 
The DAG in the lower half is generated from the quantum circuit above. 
For example, the gate $g3$ depends on gate $g1$ because qubit $q_2$ is in both $g1$ and $g3$ and $g3$ can not be executed before $g1$. 

\textbf{Front layer initialization.} 
A front layer (denoted as $F$) in this paper is defined as the set of all the two-qubit gates which have no unexecuted predecessors in the DAG.
These gates can be executed instantly from a software perspective.
For a two-qubit gate $CNOT(q_i,q_j)$, it can be placed in the set $F$ when all previous gates on $q_i$ or $q_j$ have been executed. 
By checking the generated DAG, we can select all vertices in the graph with $0$ indegree, which means the corresponding two-qubit gates have no dependencies, to initialize $F$.
In Figure~\ref{fig:DAG}, the initial front layer contains $g1$ and $g2$ because they have no predecessors.

\textbf{Temporary initial mapping generation.} 
\myalgorithmname~does not give the initial mapping at the preprocessing stage but a temporary initial mapping is still required to start our heuristic search. 
We randomly generate an initial mapping as a start point. 
Later in Section~\ref{sec:init}, we will finally update this initial mapping at the end of \myalgorithmname.

\subsection{SWAP-Based Heuristic Search}~\label{sec:flow}
The preprocessing stage leads to the distance matrix $D[~][~]$, circuit DAG, initial $F$, and an initial mapping. 
In this section, we introduce the complete SWAP-Based heuristic search procedure.

\begin{algorithm}
\SetAlgoLined
 
\KwIn{Front Layer $F$, Mapping $\pi$, Distance Matrix $D$, Circuit DAG, Chip Coupling Graph $G(V,E)$}
\KwOut{Inserted SWAPs, Final Mapping $\pi_f$}

 \While{$F$ is not empty}{
 $Execute\_gate\_list = \varnothing $ \; 
  \For{gate in $F$}{
      \If{gate can be executed on device}{
         $Execute\_gate\_list.append(gate) $\;
    }
  }
   \eIf{$Execute\_gate\_list \neq \varnothing$}{
   \For{gate in $Execute\_gate\_list$}{
        $F.remove(gate)$\;
        obtain successor gates from DAG\;
        \If{successor gates' dependencies are resolved}{
        $F.append(gate)$\;
        }
    }
    Continue\;
    }{
    $score = []$\;
    $SWAP\_candidate\_list = Obtain\_SWAPs(F, G)$\;
    \For{SWAP in $SWAP\_candidate\_list$}{
    $\pi_{temp} = \pi.update(SWAP)$\;
    $score[SWAP] = \textbf{H}(F, DAG, \pi_{temp}, D, SWAP)$\;
    }
    Find the SWAP with minimal score\;
    $\pi = \pi.update(SWAP)$\;
    }
}
\caption{\myalgorithmname's SWAP-based Heuristic Search}
\label{overall}
\end{algorithm}

 Algorithm~\ref{overall} shows the pseudo code of our search algorithm for one traversal, which scans the entire DAG from the initial front layer to the end and inserts SWAPs to make all CNOT gates executable. 
Later in Section~\ref{sec:init}, this procedure will be used multiple times 
to update the initial mapping and improve the results.
Generally, \myalgorithmname's heuristic search will iterate until the front layer $F$ is empty, which means all the gates in the circuit have been executed and the algorithm should stop.
In each iteration, it will first check if there are any gates in $F$ that can be directly executed on the chip. 
If so, it will execute these gates, remove them from $F$, and then add new gates to $F$ if possible. 
Otherwise, it will try to search for SWAPs, insert the SWAPs in the circuit, and update the mapping. 
A detailed explanation of each step  is listed as follows:

\begin{itemize}
\item Our heuristic search algorithm will first check if $F$ is empty. 
If so, all the two-qubit gates in the circuits have been executed and we should finish our search algorithm.
Otherwise, it will initialize an $Execute\_gate\_list$ and try to add some gates from $F$ to $Execute\_gate\_list$.

\item To determine whether a gate should be added into $Execute\_gate\_list$, \myalgorithmname's search algorithm will extract the logical qubits, $q_i$ and $q_j$, in the gate and use the current mapping to find the corresponding physical qubits $Q_m,Q_n = \pi(q_i), \pi(q_j)$ on the chip. 
If $Q_m$ and $Q_n$ are connected by an edge in the coupling graph $G$, then this two-qubit gate on $q_i$ and $q_j$ can be executed directly and will be added to $Execute\_gate\_list$.

\item If $Execute\_gate\_list$ is not empty, all gates in the list is removed from $F$.
After that, we will check the successor gates of these executed gates. 
For a successor CNOT gate on $q_i$ and $q_j$, if there is no gate in $F$ that is applied on any of them, then logically this successor gate is ready to be executed and we will add it to $F$. 
After executing some gates and adding the successor gates, we will go back to the beginning and the check for the executable gates again.

\item If $Execute\_gate\_list$ is empty, all the gates in $F$ 
can be executed in software but not on hardware. 
SWAPs need to be inserted to move the logical qubits in a two-qubit gate close to each other.

\item Instead of searching for a mapping, which will require exponential time and space, we only search for SWAPs associated with the qubits in $F$~(in Section~\ref{sec:searchswap}).
Suppose $q_1$  is a target of a two-qubit gate in $F$ now, we find the corresponding physical qubit $Q_i=\pi(q_1)$ in $G$ and then locate all its $5$ neighbors $Q_{i1}, \dots, Q_{i5}$. 
After that we use reverse mapping to find the corresponding logical qubits $q_{i1},\dots,q_{i5} = \pi^{-1}(Q_{i1}),\dots,\pi^{-1}(Q_{i5}),$.
For logical qubit pairs $(q_1, q_{i1}), \dots, (q_1,q_{i5})$, it is possible to insert a SWAP between the two qubits in a qubit pair since their corresponding physical qubits are connected by an edge in the coupling graph and two-qubit gates between these two qubits are supported by the hardware. 
The SWAPs on these qubit pairs will be added to $SWAP\_candidate\_list$.
We repeat the procedure above for all the qubits involved in $F$.

\item A heuristic cost function $H$ is then used to rate each SWAP in the $SWAP\_candidate\_list$.
The SWAP with the lowest score is then selected and used to update the mapping $\pi$.
After this step, the algorithm continues to check for executable gates if $F$ is not empty; otherwise, it terminates.

\end{itemize}

\begin{figure*}[!t]
\centering
\includegraphics[width=2.0\columnwidth]{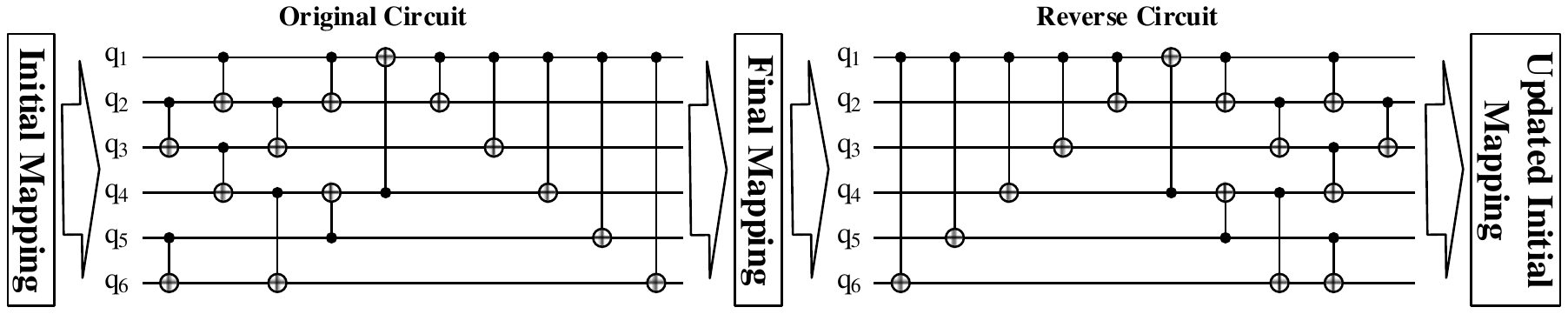}
\vspace{-5pt}
\caption{Initial Mapping Update Using Reverse Traversal Technique }
\label{fig:reverse}
\end{figure*}

\subsection{Key Design Decisions}\label{sec:decision}
Compared with previous solutions, \myalgorithmname~features three points to ensure the design objectives can be achieved. Three corresponding examples are given to demonstrate the benefits of our design decisions.

\subsubsection{SWAP-Based Search Scheme}\label{sec:searchswap}
Previous works usually employ mapping-based exhaustive search to find the valid mapping transition with low overhead~\cite{wille2016look,zulehner2018efficient}. 
For example, Zulehner \textit{et al.} search all possible combination of SWAPs that can be applied concurrently trying to minimize the output circuit depth and the number of additional SWAPs simultaneously~\cite{zulehner2018efficient}. 
However, such exhaustive search requires $O(exp(N))$ time and space, which makes the algorithms not applicable to larger-size NISQ devices~(experimental results discussed in Section~\ref{sec:limitofbestknown}).

\begin{figure}[!b]
\centering
\includegraphics[width=1.0\columnwidth]{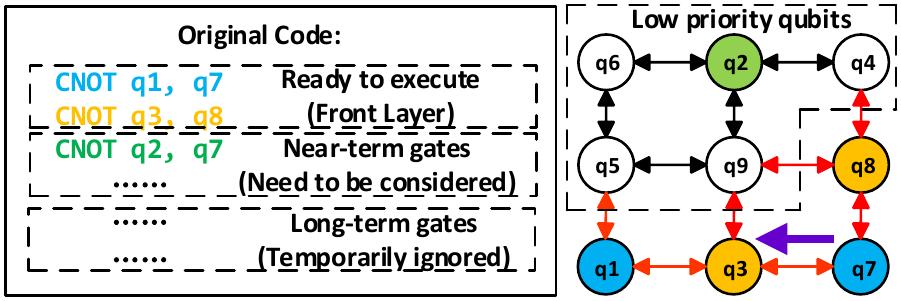}
\vspace{-15pt}
\caption{Example of SWAP-Based Heuristic Search}
\label{fig:searchswap}
\end{figure}

We observe that many SWAPs in the mapping-based exhaustive search can be redundant and eliminated. Figure~\ref{fig:searchswap} shows an example of how we reduce the search space and find the SWAP. 
Suppose we have a 9-qubit device. The coupling graph and initial mapping are shown on the right side. 
The program we need to execute is on the left side.
The first two CNOT gates are in the front layer and ready to be executed.
The third CNOT needs to be executed after the first one due to the dependency on $q_{7}$. 
The first two gates cannot be executed directly because their corresponding physical qubit pairs are not connected. 
All qubits not involved in the front layer~($q_{2}$, $q_{4}$, $q_{5}$, $q_{6}$, $q_{9}$) are considered as low priority ones and any SWAPs inside this low priority qubit set cannot help with resolving dependencies in the front layer. 
Thus, only the SWAPs that associate with at least one qubit in the front layer~(the edges marked red in Figure~\ref{fig:searchswap}) are the candidate SWAPs .

For all the candidate SWAPs, we design a heuristic cost function to help find the SWAP that can reduce the sum of distances between each qubit pairs in the front layer. 
Moreover, we also enable look-ahead ability in the heuristic cost function by considering the gates right after the front layer. 
The detailed design of our heuristic cost function is in Section~\ref{sec:heuristic}.
Here in this example in Figure~\ref{fig:searchswap}, we can find that the SWAP marked by a purple arrow is the best one. 
It can make all the CNOT gates in the front layer executable and also reduce the distance between $q_{2}$ and $q_{7}$, which are in a CNOT gate right after the front layer.
For the long-term gates far away from the front layer, we temporarily do not consider them because the mapping may vary a lot during execution and it is hard to estimate the cost accurately over a long gate sequence without exhaustive search.

\textbf{Complexity Analysis.} An upper bound of the computation complexity can be estimated by the worst case, in which each two-qubit gate is satisfied individually. 
The problem finishes when all two-qubit gates have been satisfied.
The time complexity to satisfy one two-qubit gate is the multiplication of the time to evaluate a potential option in the search space, the size of the largest possible search space, and the maximum number of search steps per two-qubit gate.
The complexity of the heuristic cost function computation is $O(N)$~(in Section~\ref{sec:heuristic}). 
This SWAP-based search could bring exponential speedup by reducing the search space from $O(exp(N))$ to $O(N)$ (in the worst case all the qubits are involved in the front layer), which makes \myalgorithmname~scalable to larger size cases. 
Although it increases the number of search steps because we may need multiple SWAPs for one two-qubit gate, the benefit is still significant because we need at most the diameter of the chip coupling graph~($O(\sqrt{N})$ for 2D layout) number of SWAPs to move two qubits together for each two-qubit gate. 
In summary, our SWAP-based search scheme can reduce the complexity from $O(exp(N))$ to at most $O(N^{2.5})$
for each two-qubit gate, which makes \myalgorithmname~exponentially faster as $N$ increases.

\subsubsection{Reverse Traversal for Initial Mapping}~\label{sec:init}
It has been proved that initial mapping could have a huge impact on the final result~\cite{siraichi2018qubit,zulehner2018efficient}.
However, no previous solution could give an initial mapping with global consideration. 
Siraichi \textit{et al.} counted the number of coupled logical qubits in the circuit for each logical qubit and tried to find a match with the outdegree of the physical qubit in the coupling graph with no temporal information considered~\cite{siraichi2018qubit}.
Zulehner \textit{et al.} determined the initial mapping by those two-qubit gates at the beginning of the circuit without global consideration~\cite{zulehner2018efficient}.

Different from classical circuit or programs, quantum circuits are reversible. You can easily generate a reverse circuit of the original circuit. 
The two-qubit gates in the reverse circuit will be exactly the same with only the order reversed. 
Figure~\ref{fig:reverse} shows an example of the reverse circuit. 
The last (first) CNOT gate in the original circuit will be the first (last) CNOT gate in the reverse circuit on the same qubits.
This symmetry between the original circuit and the reverse circuit brings us a new opportunity for initial mapping optimization.
If we know the final mapping of a quantum circuit, we can use this final mapping as the initial mapping to solve qubit mapping problem for the reverse circuit on the same hardware model. 
The final mapping of the reverse circuit can be an initial mapping for the original circuit. 
This updated initial mapping comes with better quality because all the gates' information is considered. The gates that are closer to the beginning of the circuit will have more impact on the initial mapping optimization. 
The gates far away from the beginning have less impact but can still be considered through these forward and backward traversals.

Based on this observation, we propose a novel reverse traversal technique to generate high-quality initial mapping with global information considered. 
Figure~\ref{fig:reverse} illustrates the procedure of this technique. 
We first randomly generate an initial mapping and then apply our SWAP-based heuristic search to traverse through the original circuit. 
The final mapping obtained from this forward traversal will be used as the initial mapping in the following reverse traversal. 
We use the same SWAP-based search with only the circuit reversed and the original initial mapping will be updated to the final mapping in the reverse traversal.

\subsubsection{Trade-off between the Circuit Depth and the Number of Gates}\label{sec:tradeoff}
\begin{figure}[!t]
\centering
\includegraphics[width=1.0\columnwidth]{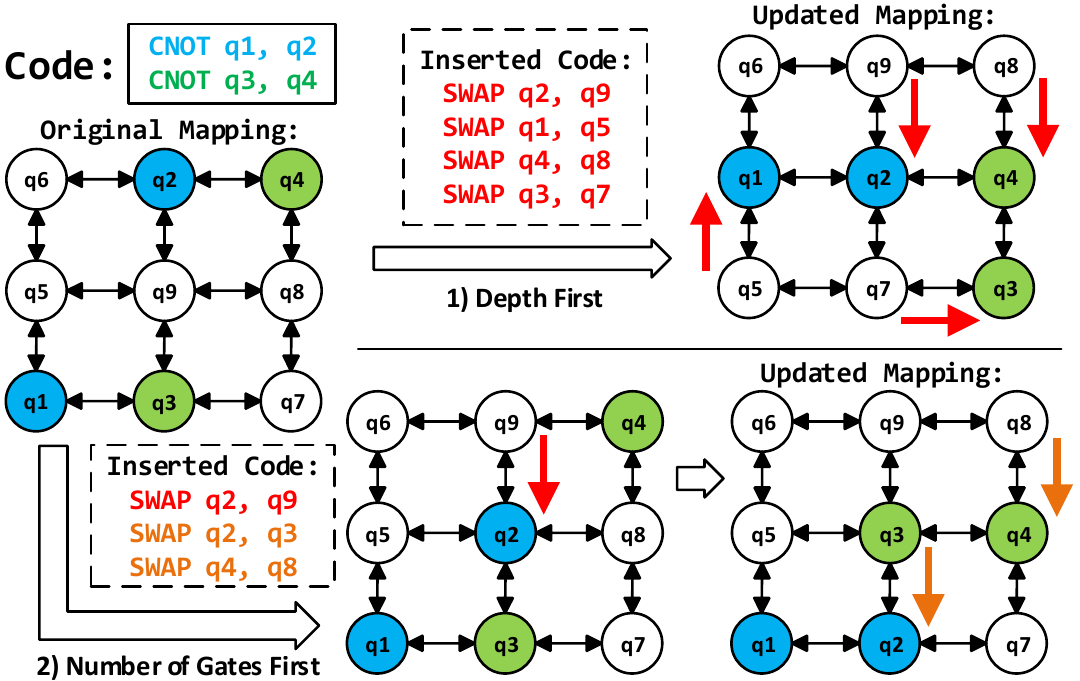}
\vspace{-15pt}
\caption{Example of Generated Circuits for Different Optimization Objectives~(a Trade-off between $d$ and $g$)}
\vspace{-5pt}
\label{fig:tradeoffexmaple}
\end{figure}

When we insert SWAPs in the original quantum circuit, there is a trade-off between these two metrics, the number of gates and the circuit depth~(an analogy in classical computation can be the trade-off between area and latency in digital circuit design). 
Figure~\ref{fig:tradeoffexmaple} shows an example.
Suppose there is a 9-qubit device and we have 2 CNOT gates on $\{q1,q2\}$, $\{q3,q4\}$~(marked by blue and green) to execute.
The initial mapping is shown on the left side. We have two different solutions with different optimization objectives: 
\textbf{1) Depth First}. 
By inserting 4 non-overlap SWAPs on $\{q1,q5\}$, $\{q2,q9\}$ $\{q3,q7\}$, and $\{q4,q8\}$~(marked by 4 red arrows) which can be executed simultaneously, we can satisfy these 2 two-qubit gate dependencies with 4 additional SWAPs and the circuit depth increases by 1 SWAP. 
\textbf{2) Number of Gates First}. 
$\{q2,q9\}$ is first swapped and then two qubit pairs $\{q2,q3\}$ and $\{q4,q8\}$ are swapped simultaneously. 
The SWAP on $\{q2,q3\}$ must be applied after the first SWAP on $\{q2,q9\}$ so that the circuit depth  increases by 2 SWAPs but only 3 additional SWAPs are required to resolve all the dependencies.

The two solutions above showed an example of a trade-off between $d$ and $g$. 
However, no previous solutions can exploit this trade-off and control the generated circuit among different design optimization. A $decay$ effect is introduced in \myalgorithmname~which makes our heuristic search algorithm tend to select non-overlap SWAPs. For example, after the SWAP on $q_2$ and $q_9$, the heuristic cost function result for any SWAPs containing $q_2$ or $q_9$ will slightly increase to let our search algorithm tend to choose SWAPs containing other qubits.

In summary, these three design decisions above bring us exponential speedup for scalability, an initial mapping solution with high-quality, and the controllability between different optimization objectives. These advantages make \myalgorithmname~achieve all the design objectives discussed in Section~\ref{sec:objective}.

\subsection{Design the Heuristic Cost Function}\label{sec:heuristic}
As mentioned above, the objectives for heuristic cost function are summarized as follows:
\begin{enumerate}
\item $H$ should be able to indicate the SWAP that can move the qubits in $F$ closer to finally allow the physical execution of the two-qubit gates in $F$.
\item Besides the two-qubit gates in $F$, the heuristic cost function should be able to consider more following two-qubit gates to enable more effective qubit movement.
\item It should be able to control the parallelism of inserted SWAPs to enable the trade-off mentioned in Section~\ref{sec:tradeoff}.
\end{enumerate}

Nearest Neighbor Cost~(NNC) function is used to construct the basic heuristic function. Further optimization is introduced later to achieve all the design objectives.

\subsubsection{Nearest Neighbor Cost Function.}
NNC-based heuristic function has been widely used in previous research~\cite{wille2016look,zulehner2018efficient,saeedi2011synthesis}. 
NNC is the minimal number of SWAPs required to move two logical qubits adjacent to each other on the quantum device. 
On ideal 1D/2D lattice hardware models, NNC can be easily obtained from the coordinates of the physical qubits while on NISQ devices with irregular coupling, NNC is the length of the shortest path between two physical qubits on the coupling graph, which 
has already been obtained in $D[~][~]$ during the preprocessing stage~(an offset -1 is ignored without affecting the result).
In our design, the summation of the distances between all qubit pairs in $F$ is the basic heuristic cost function~(shown in Equation~\ref{eq:basic}). 
To evaluate the candidate SWAPs, the mapping $\pi$ is temporarily changed by a SWAP and then $H_{basic}$ is calculated.
If $H_{basic}$ is small, it means generally the distances between the two qubits in the qubit pairs from $F$ are short and this SWAP is more likely to make the gates in $F$ executable.
The SWAP with the minimal $H_{basic}$ will be selected. 

\begin{equation}\label{eq:basic}
\begin{aligned}
H_{basic} = \sum_{gate \in F}D[\pi(gate.q_1)][\pi(gate.q_2)] 
\end{aligned}
\end{equation}

\subsubsection{Look-Ahead Ability and Parallelism.}~\label{sec:optimizeH}
Although $H_{basic}$ is able to guide the heuristic search and solve the qubit movement, it only considers the two-qubit gates in $F$. 
However, a local qubit movement can affect not only the gates in $F$ but also the following gates. For the example in Figure~\ref{fig:swap_example}, the SWAP between $q_3$ and $q_7$ is a good selection because it not only resolves the dependencies for the gates in the front layer but also makes the $q_2$ and $q_7$ closer in the following gate.
Thus, we introduce the Extended Set $E$, which contains some closet successors of the gates from $F$ in the DAG.
The size of $E$ is flexible, depending on how much look-ahead ability we hope to have.
A large $E$ is not necessary since the summation over $E$ is only an inaccurate estimation of the effect of a SWAP and the amount of computation will also increase.

In the updated heuristic cost function, we sum over the gates in both $E$ and $F$ to enable the look-ahead ability.
Since $E$ and $F$ has different sizes, we normalize the two summations by the sizes of $F$ and $E$ respectively.
Also, the gates in $F$ should have some priority since they need to be executed before those in $E$. 
So that a weight parameter $W,  0 \leq W < 1$, is added to reduce the effect of the second term.

\begin{table*}[!t]
  \centering
  \caption{Number of Additional Gates and Runtime Compared with BKA~\cite{zulehner2018efficient}}
  \resizebox{\textwidth}{!}{
    \begin{tabular}{|c|c|c|c|c|c|c|c|c|c|c|c|c|c|}

    \hline
    \multicolumn{4}{|c|}{Original Circuit} & \multicolumn{3}{c|}{BKA~\cite{zulehner2018efficient}~(C++)} & \multicolumn{4}{c|}{\myalgorithmname~(Python)} & \multicolumn{3}{c|}{Comparison} \\
     \hline
    type & name & n & $g_{ori}$ & $g_{add}$ & $g_{tot}$    & $t_{tot}$     &$g_{la}$& $g_{op}$ & $t_1$& $t_{op}$ & $t_{tot}/t_{op} $  & $\Delta g$ & $\Delta g/g_{add}$ \\
    \hline
        
 \hline
   small & 4mod5-v1\_22 & 5     & 21    & 15    & 36    & 0     & 6     & 0     & 0     & 0     &   N/A    & 15    & 100\% \\
     \hline
   small &  mod5mils\_65 & 5     & 35    & 18    & 53    & 0     & 12    & 0     & 0     & 0     &   N/A    & 18    & 100\% \\
  \hline
    small &  alu-v0\_27 & 5     & 36    & 33    & 69    & 0     & 30    & 3     & 0     & 0     &   N/A    & 30    & 91\% \\
 \hline
  small &   decod24-v2\_43 & 4     & 52    & 27    & 79    & 0     & 9     & 0     & 0     & 0     &  N/A     & 27    & 100\% \\
  \hline
    small &  4gt13\_92 & 5     & 66    & 42    & 108   & 0     & 18    & 0     & 0     & 0     &    N/A   & 42    & 100\% \\
 \hline
 
 \hline
  sim&  ising\_model\_10 & 10    & 480   & 18    & 498   & 1.37  & 39    & 0     & 0.003 & 0.004 & 342.5 & 18    & 100\% \\
 \hline
  sim&  ising\_model\_13 & 13    & 633   & 60    & 693   & 42.46 & 66    & 0     & 0.005 & 0.007 & 6066 & 60    & 100\% \\
 \hline
   sim& ising\_model\_16 & 16    & 786   & \multicolumn{3}{c|}{Out of Memory} & 84    & 0     & 0.008 & 0.01  &   N/A    &  N/A     &  N/A\\
 \hline
 
 \hline
   qft& qft\_10 & 10    & 200   & 66    & 266   & 0.22  & 93    & 54    & 0.004 & 0.103 & 2.136& 12    & 18\% \\
    \hline
  qft&  qft\_13 & 13    & 403   & 177   & 580   & 266.27 & 204   & 93    & 0.015 & 0.036 & 7396 & 84    & 47\% \\
    \hline
  qft&  qft\_16 & 16    & 512   & 267   & 779   & 474.81 & 276   & 186   & 0.028 & 0.084 & 5652 & 81    & 30\% \\
    \hline
  qft&  qft\_20 & 20    & 970   & \multicolumn{3}{c|}{Out of Memory} & 429   & 372   & 0.034 & 0.102 &    N/A   &   N/A    &  N/A\\
    \hline
    
 \hline
 large&   rd84\_142 & 15    & 343   & 138   & 481   & 1.97  & 243   & 105   & 0.012 & 0.035 & 56.29 & 33    & 24\% \\
    \hline
  large&   adr4\_197 & 13    & 3439  & 1722  & 5161  & 4.53  & 2112  & 1614  & 0.19  & 0.49  & 9.245 & 108   & 6\% \\
    \hline
  large&   radd\_250 & 13    & 3213  & 1434  & 4647  & 2.23  & 1488  & 1275  & 0.16  & 0.48  & 4.646 & 159   & 11\% \\
    \hline
   large&  z4\_268 & 11    & 3073  & 1383  & 4456  & 1.15  & 1695  & 1365  & 0.15  & 0.44  & 2.614 & 18    & 1\% \\
    \hline
    large& sym6\_145 & 14    & 3888  & 1806  & 5694  & 0.56  & 1650  & 1272  & 0.19  & 0.56  & 1.000     & 534   & 30\% \\
    \hline
   large&  misex1\_241 & 15    & 4813  & 2097  & 6910  & 0.3   & 2904  & 1521  & 0.29  & 0.89  & 0.337 & 576   & 27\% \\
    \hline
   large&  rd73\_252 & 10    & 5321  & 2160  & 7481  & 1.19  & 2391  & 2133  & 0.31  & 0.94  & 1.266 & 27    & 1\% \\
    \hline
    large& cycle10\_2\_110 & 12    & 6050  & 2802  & 8852  & 1.31  & 2622  & 2622  & 0.44  & 1.35  & 0.970 & 180   & 6\% \\
    \hline
   large&  square\_root\_7 & 15    & 7630  & 3132  & 10762 & 2.81  & 5049  & 2598  & 0.63  & 1.5   & 1.873& 534   & 17\% \\
    \hline
   large&  sqn\_258 & 10    & 10223 & 4737  & 14960 & 16.92 & 5934  & 4344  & 1.23  & 3.52  & 4.807 & 393   & 8\% \\
    \hline
   large&  rd84\_253 & 12    & 13658 & 6483  & 20141 & 15.25 & 7668  & 6147  & 1.82  & 5.39  & 2.829 & 336   & 5\% \\
    \hline
 large&    co14\_215 & 15    & 17936 & 9183  & 27119 & 18.37 & 10128 & 8982  & 3.18  & 9.51  & 1.932 & 201   & 2\% \\
    \hline
  large&   sym9\_193 & 10    & 34881 & 17496 & 52377 & 72.61 & 26355 & 16653 & 11.11 & 30.17 & 2.407 & 843   & 5\% \\
  \hline
   large&  9symml\_195 & 11    & 34881 & 17496 & 52377 & 81.73 & 25368 & 17268 & 11.1  & 31.42 & 2.601 & 228   & 1\% \\
    \hline
    \end{tabular}
}%
    
    \begin{scriptsize}
    \begin{tablenotes}
\item 
\textbf{small:} small quantum arithmetic. 
\textbf{sim:} quantum simulation. 
\textbf{qft:} quantum fourier transform. 
\textbf{large:} large quantum arithmetic. 
\textbf{$n$:} number of logical qubits in the original circuit. 
\textbf{$g_{ori}$:} original number of gates. 
\textbf{$g_{add}$:} number of additional gates. 
\textbf{$g_{tot}$:} total number of gates. 
\textbf{$t_{tot}$:} total runtime in seconds, `0' means shorter than 0.001 second. 
\textbf{$g_{la}$:} number of additional gates with only look-ahead heuristic. 
\textbf{$g_{op}$:} number of additional gates after reversal traverse. 
\textbf{$t_{1}$:} runtime of first traverse in seconds. 
\textbf{$t_{op}$:} runtime of all 3 traversals. 
\textbf{$\Delta g$:} $=g_{add}-g_{op}$. 
\textbf{Out of Memory:} the program required more than 378 GB memory~(entire memory space on the test server) 
 \end{tablenotes}
 \end{scriptsize}
    
  \label{tab:result}%
\end{table*}%

In order to select SWAPs that can be executed in parallel, a $decay$ effect is introduced in the heuristic cost function. 
If a qubit $q_{i}$ is involved in a SWAP recently, then its $decay$ parameter will increase by $\delta$~($decay(q_{i}) = 1 + \delta$).
This decay parameter will let our heuristic search tend to select non-overlap SWAPs and increase the parallelism in the generated circuit. 
Moreover, by tuning the value of $\delta$, we are able to control the `willingness' of our heuristic search to generate different circuits with a trade-off between the number of gates and circuit depth. 
The final version of our optimized heuristic function is shown in Equation~\ref{eq:Hopt}.
The complexity of this heuristic function is $O(N)$ since all qubits appear in $F$ in the worst case. 
The size of $E$ is not considered because it will not be very large and is set to $N$ in our evaluation.

\begin{equation}~\label{eq:Hopt}
\begin{aligned}
H = max(decay(SWAP.q_1), decay(SWAP.q_2)) \\
*\{\frac{1}{|F|}\sum_{gate \in F}D[\pi(gate.q_1)][\pi(gate.q_2)] \\
+ W * \frac{1}{|E|}\sum_{gate \in E}D[\pi(gate.q_1)][\pi(gate.q_2)]\}
\end{aligned}
\end{equation}

\section{Evaluation}\label{sec:evaluation}
In this section, we evaluate \myalgorithmname~with a set of benchmarks on the latest, reported hardware model based on the superconducting circuit technology.

\textbf{Benchmarks.} 
The benchmarks are selected from previous work~\cite{siraichi2018qubit,zulehner2018efficient}, including quantum programs from IBM's QISKit~\cite{IBMqiskit}, some functions from RevLib~\cite{wille2008revlib}, and some algorithms compiled from Quipper~\cite{green2013quipper} and ScaffCC~\cite{javadiabhari2014scaffcc}. 

\textbf{Hardware Model.} 
We use the coupling graph from IBM's latest Q 20 Tokyo chip~\cite{IBMdevice}~(Figure~\ref{fig:ibmq20}). 
All the couplings are symmetric and the CNOT gate is allowed in both directions between each pair of connected physical qubits.

\textbf{Experiment Platform.} All experiments in this paper are executed on a server with 2 Intel Xeon E5-2680 CPUs~(48 logical cores) and 378GB memory. The Operating System is CentOS 7.5 with Linux kernel version of 3.10.

\textbf{Algorithm Configuration.} The size of the Extended Set $|E|$ is fixed to be 20 and the the weight $W$ to be $0.5$. The $decay$ parameter $\delta$ increases from $0.001$ and this $decay$ function is reset every 5 search steps or after a CNOT gate is executed.
The algorithm is executed for 5 times, each with a different initial mapping for each benchmark. Each time we run 3 traversals~(forward-backward-forward) and report the best result out of 5 attempts.

\textbf{Comparison.} There are several existing algorithms with the flexibility to be applied to an arbitrary coupling graph proposed by IBM~\cite{IBMqiskit}, Siraichi \textit{et al.}~\cite{siraichi2018qubit}, and Zulehner \textit{et al.} ~\cite{zulehner2018efficient}.
Among them, Zulehner \textit{et al.}'s algorithm has beaten the other two solutions and is used as the Best Known Algorithm~(BKA) in this paper.
For a fair comparison, their source code~\cite{robertsolution} is downloaded and only the embedded hardware model is modified to be the same IBM 20-qubit chip model. It is then recompiled with full optimization, and executed on the same server with \myalgorithmname.

\subsection{Number of Gates Reduction}
Table~\ref{tab:result} shows the gate counts reduction of \myalgorithmname~compared with BKA~\cite{zulehner2018efficient}.
\myalgorithmname~could beat BKA on various benchmarks of different sizes.

\subsubsection{Small Size Cases and Ising Model} \myalgorithmname~could perform much better than BKA on small-size benchmarks.
It is able to find a good initial qubit mapping with no or very few additional SWAPs required. 
The number of additional gates could be significantly reduced by 91\% or even fully eliminated. 
For ising model benchmarks, the optimal solution is trivial since the ising model in quantum mechanics only considers nearby coupling energy. 
Although the number of qubits and the number of gates are much larger compared with small cases, \myalgorithmname~can still find the optimal solution. 
BKA only considers the two-qubit gates at the beginning of the circuit without such a scheme to improve the initial mapping. 

\subsubsection{Large Size Cases} For larger circuits in type `large' and `qft', \myalgorithmname~can still be better than BKA. 
Since the BKA searches a much larger space in each step, \myalgorithmname~may not achieve the same or better result in the first traversal.
The $g_{la}$ column in Table~\ref{tab:result} shows the number of additional gates after the first traversal with look-ahead heuristic function and $g_{la}$ is larger than $g_{add}$ in most cases. 
However, with the help of our reverse traversal technique, \myalgorithmname~(shown in $g_{re}$) is able to outperform BKA with the updated initial mapping and reduce the number of additional gates by $10\%$ on average. 

Note that the gate count reduction for large size cases is less significant than that for small size cases. 
This difference comes from whether a perfect initial mapping, which could satisfy all the CNOT gate constraints in the program after the inital mapping and does not require further SWAPs, can be found.
For small benchmarks, there often exists a physical qubit coupling subgraph that can perfectly or almost match logical qubit coupling in the benchmarks. 
Our algorithm can find such matching (at least for all small benchmarks we have tested), while BKA cannot. This leads to substantial benefit since very few or no SWAPs are inserted. 
For the benchmarks with larger number of gates, a physical qubit subgraph that can match the logical qubits coupling usually does not exist. Therefore, both our approach and the baseline need to insert more SWAPs, leading to less benefit.

\subsection{Runtime Speedup and Scalability}
As discussed in Section~\ref{sec:searchswap}, the size of search space is $O(exp(N))$ in BKA, which limits its scalability in terms of the number of qubits. 
But the search space size in \myalgorithmname~is only $O(N)$. 
Although more search steps are required since only one SWAP is selected in each step, the overall complexity in the worst case is still $O(N^{2.5}g)$.
Such a difference in complexity makes BKA not applicable to larger size cases.

\subsubsection{Runtime Comparison} BKA is written in  C++ and compiled with GCC O3 optimization, while \myalgorithmname~is implemented in pure Python without any parallelization or C/C++ accelerated library. 
The `$t_{tot}/t_{op}$' column in Table~\ref{tab:result} shows the ratio between the execution time of BKA and \myalgorithmname. For most benchmarks, \myalgorithmname~requires much less execution time. 
Even in the worst case `misex1\_241', \myalgorithmname~only needs about 3 times runtime compared with the BKA. 
Since the intrinsic speed difference between C++ and Python can be over 100 times, the speedup can still be estimated to be dozens of times 
if the same programming language is used.

\subsubsection{Limit of BKA and Scalability}~\label{sec:limitofbestknown} Our experiments have reached the limit of BKA~(shown in Table~\ref{tab:result} with `Out of Memory').
For the `sim' and `qft' type, the benchmarks share the same function with different input sizes.
The runtime of BKA grows rapidly as the number of qubits increases. 
For `qft\_16' benchmark, we observe that BKA requires more than \textbf{40GB} memory and \textbf{474.81} seconds runtime while \myalgorithmname~only required about \textbf{200MB} memory and \textbf{0.08} seconds runtime.
For `ising\_model\_16' and `qft\_20' benchmarks, the BKA requires more than \textbf{378GB} memory and can not be executed on our server. 
But \myalgorithmname~can still solve it in \textbf{0.1} seconds with about \textbf{300MB} memory. 
These results show that \myalgorithmname~is much more scalable than BKA.

\subsection{Trade-off between Number of Gates and Depth}\label{sec:tradeoffresult}
\begin{figure}[!ht]
\centering
\vspace{-8pt}
\includegraphics[width=1.0\columnwidth]{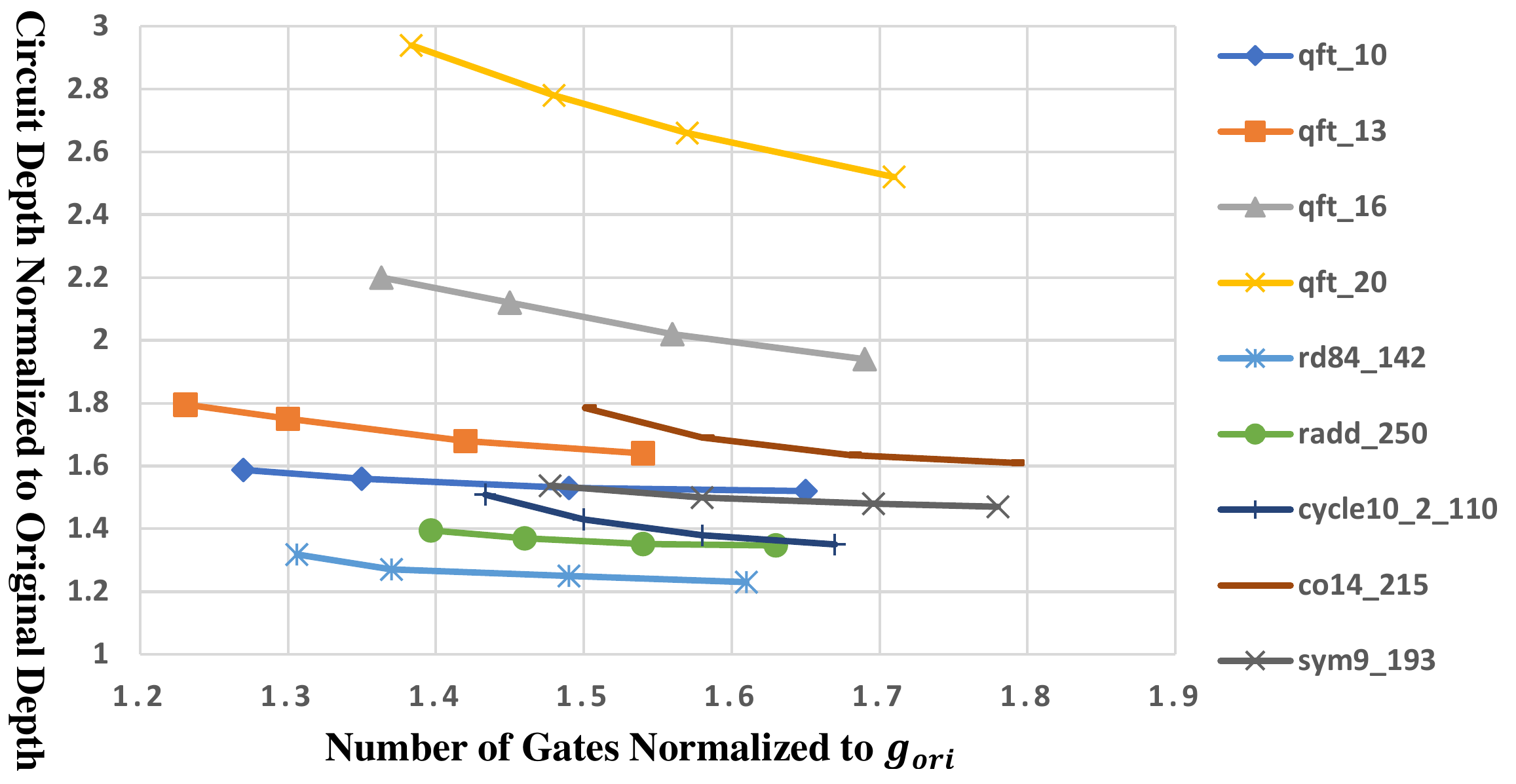}
\vspace{-20pt}
\caption{Trade-off between $g$ and $d$ in the Output Circuits}
\vspace{-15pt}
\label{fig:tradeoff}
\end{figure}

The $decay$ effect is introduced in the heuristic cost function in order to reduce the depth of the generated circuit. 
Figure~\ref{fig:tradeoff} shows the generated circuit variation with different $\delta$ values for 9 benchmarks. 
The X-axis is the number of gates normalized to $g_{ori}$~(in Table~\ref{tab:result}). 
The Y-axis represents the generated circuit depth normalized to the original circuit depth. 
These results showed that \myalgorithmname~could provide about $8\%$ variation in generated circuit depth by varying the number of gates and control the generated circuit quality. 
For a specific implementation technology, we can change the $\delta$ according to the qubit coherence time and gate fidelity data.
However, if we continue to increase $\delta$, both the circuit depth and the number of gates may increase~(not shown in the figure) because our search algorithm will consider more about unmoved qubits instead of trying to satisfy a CNOT dependency, which will bring redundant SWAPs.

\section{Limitation and Future Work}\label{sec:futurework}
This paper provides an effective, flexible, and scalable solution for the qubit mapping problem. 
However, some of our assumptions may not hold due to the rapid development in this area. Some limitations and potential future research directions are listed as follows:

\textbf{Benchmarks.}
We select 26 benchmarks of different sizes and functions from several benchmark suites. 
However, these quantum circuits may not be able to fully represent the characteristics of emerging practical NISQ applications which are still under development.

\textbf{Different Chip Architecture.} We use the hardware model from the latest IBM's 20-qubit chip, on which each connected qubit pair support CNOT gates. 
However, the chip model varies among different vendors. For example, Rigetti's QPU supports CPhase and iSWAP two-qubit gates~\cite{RigettiQPU,sete2016functional}. 
How to design more general circuit transformations is beyond the scope of this paper but can be a future research direction.


\textbf{More Precise Hardware Modeling.} Besides the qubit coherence time, gate fidelity, and available on-chip coupling, the difference in the error rate of various quantum gates and of the same quantum gate applied on different qubits or qubit pairs may also influence the fidelity of executing a quantum algorithm~\cite{tannu2018case}. 
In addition, realistic hardware suffers from more imperfections which are not covered in this paper, such as the cross talk between qubits. Both facts call for a more precise hardware model to enable better platform-specific quantum circuit optimization.

\section{Related Work}\label{sec:relatedwork}

Although the qubit mapping problem shares some similarities with the register allocation~\cite{chaitin1982register,poletto1999linear} and instruction scheduling problem~\cite{tomasulo1967efficient,hennessy1983postpass,codina2001unified} in classical computing, the constraints are completely different. 
In register allocation, the main constraint is the limited number of registers while for quantum computing, the number of physical qubits cannot be smaller than that of logical qubits.
In instruction scheduling, the main constraints are the data dependency and limited number of computing units but in the qubit mapping problem, the major constraint is the limited coupling between physical qubits.
Therefore, existing methods for such problems cannot be directly applied in this qubit mapping problem.

It is well known that nearest neighbor coupling is the most feasible and promising when there were only devices with a very limited number of qubits.
Attempts to solve qubit mapping problem at that time were made on hypothetical quantum hardware models like ideal 1D/2D lattice models and can be classified into two types.
One popular type of approach is to formulate the qubit mapping problem into a mathematically equivalent optimization problem and then apply a software solver~\cite{maslov2008quantum,chakrabarti2011linear,shafaei2013optimization,shafaei2014qubit,wille2014optimal,lye2015determining,venturelli2017temporal,venturelli2018compiling,booth2018comparing,oddi2018greedy,bhattacharjee2017depth}.
The major drawback of this type of approach is that a general solver cannot utilize the intrinsic feature in qubit mapping and the execution time is usually very long compared with the following heuristic approaches.
Another type of approach is search algorithms guided by heuristic cost functions.
Several attempts have been made on ideal 1D/2D lattice qubit coupling models~\cite{alfailakawi2014lnn,saeedi2011synthesis,lin2015paqcs,wille2016look,shrivastwa2015fast,kole2016heuristic,kole2018new,bhattacharjee2018novel}, but they are not applicable in the NISQ era since qubit coupling can be much more complex and restricted on NISQ devices.
Some other works target hypothetical large-scale quantum computers~\cite{heckey2015compiler,javadi2017optimized}, which is beyond the scope of NISQ and the qubit mapping problem turns out to be another one~\cite{heckey2015compiler,paler2017fault,javadi2017optimized,lao2018mapping}.

After IBM launched its quantum cloud service, more people could work on hardware models from realistic devices.
IBM provides a mapper targeting IBM's chips in its quantum computing toolkit QISKit~\cite{IBMqiskit}. 
This mapper divides the quantum circuit into independent layers. Each layer only contains non-overlapped operations. Then it randomly searches satisfying mappings for each layer guided by certain heuristics~\cite{siraichi2018qubit,zulehner2018efficient}.
Bisides IBM's solution, two more recent works~\cite{siraichi2018qubit,zulehner2018efficient} are proposed targeting IBM's chips and can handle devices with arbitrary coupling, which are discussed as follows.

Siraichi \textit{et al.} studied the qubit allocation problem on IBM QX2 and QX3 chips~\cite{siraichi2018qubit}.
They proposed a search algorithm to find the optimal solution based on dynamic programming.
However, this optimal algorithm requires exponential time and space to execute and can only work for circuits with 8 or fewer qubits.
For larger size cases, they proposed a heuristic method for both initial mapping and intermediate qubit movement.
Their initial mapping solution counted the number of two-qubit gates between each pair of logical qubits and tried to find a matched edge on the physical chip with no temporal information considered in this stage.
For the qubit movement, they only resolved one two-qubit gate each time and determined whether to move qubits depending on the number of two-qubit gates between them greedily without considering the effects of these local decisions. 
Their heuristic method is fast but oversimplified with results worse than IBM's solution.

Zulehner \textit{et al.} tried to use A* search plus heuristic cost function~\cite{zulehner2018efficient}~(BKA in this paper).
They divided the two-qubit gates into independent layers similar to IBM's solution. 
Then they searched all possible combination of SWAP gates to minimize the sum of distance between the coupled qubits in the layer and reduce the depth of the final output circuit at the same time.
Although their method is much faster and better than IBM's approach and only requires up to several minutes on 16-qubit circuits, searching all possible combinations of concurrent SWAP gates still requires exponential time. 
Their initial mapping was determined by only those two-qubit gates at the beginning of the circuit without global consideration.

\section{Conclusion}\label{sec:conclusion}

The NISQ era is coming in the next few years while a significant gap remains between quantum software and imperfect NISQ hardware.
This paper tried to solve the qubit mapping problem caused by limited physical qubits coupling on NISQ devices.
Two-qubit gate is allowed between arbitrary two logical qubits but can only be implemented between two nearby physical qubits on NISQ hardware.
The initial mapping between logical qubits and physical qubits and its evolution need to be carefully designed to minimize the circuit transformation overhead. 
We propose \myalgorithmname, a novel SWAP-based bidirectional heuristic search method to overcome the drawbacks of previous works and ensure flexibility, scalability, controllability, and high-quality initial mapping. 
Experiment results show that \myalgorithmname~can generate hardware-compliant circuit among different objectives with less or comparable overhead consuming much shorter execution time.
Although \myalgorithmname~works for IBM chips with arbitrary symmetric CNOT coupling, the hardware model, which differs among vendors and may change over time, 
is also simplified and single-qubit gates are not yet considered. We only add additional gates instead of modifying the original circuit, while the latter one is much more complicated.
In conclusion, this work explored one step in mitigating the quantum software-hardware gap. Future work is required to take more precise hardware models into consideration.

\bibliographystyle{IEEETran}


\end{document}